\documentclass[12pt]{article}
\begin{document}
\markright{Exotic Kaluza--Klein... \hfil 
Physics Letters {\bf B159}, 22-25 (1985).}
\title{An exotic class of Kaluza--Klein models}
\author{
Matt Visser\\[2mm]
{\small \it Physics Department,}\\
{\small \it University of Southern California,}\\
{\small \it Los Angeles, CA  90080-0484,}\\
{\small \it USA}}
\date{{\small 12 September 1985; \LaTeX-ed \today }}
\maketitle
\thispagestyle{empty}
\begin{abstract}
We discuss an exotic class of Kaluza--Klein models in which the
internal space is neither compact nor even of finite volume.  Rather
than using the usual compact internal space we consider the case where
particles are gravitationally trapped near a four-dimensional
submanifold of the higher dimensional spacetime.  A specific model
exhibiting this phenomenon is constructed in five dimensions.

\smallskip

\centerline{Physics Letters {\bf B159}, 22--25 (1985).}

\smallskip

{\sf Note:} {\it This rather old paper has recently been subject of
renewed interest due to the explosion of activity in the ``non-compact
extra dimensions'' variant of the Kaluza--Klein model. It is not
available elsewhere on the Internet, and in the interests of easy
access I am placing it on} {\sf hep-th.} {\it The body of the paper is
identical to the published version. A small separate note has been
added at the end of the paper.}
\end{abstract} 

\smallskip

\centerline{\underline{Present address:}} 
\centerline{\small \it Physics Dept, 
Washington University, St Louis, MO 63130--4899}

\medskip

\centerline{\underline{E-mail:} {\sf visser@kiwi.wustl.edu}}

\medskip

\centerline{\underline{Homepage:}
{\sf http://www.physics.wustl.edu/\~{}visser}}

\medskip

\centerline{\underline{Archive:}
{\sf hep-th/9910093}}

\def\Box{\nabla^2}
\def\d{{\mathrm d}}
\def\x{{\vec x\,}}
\def\p{{\vec p\,}}
\def\k{{\vec k\,}}
\def\half{{1\over2}}
\def\quarter{{1\over4}}
\def\L{{\cal L}}
\def\E{{\cal E}}
\def\sech{\hbox{sech}}
\def\ie{{\em i.e.}}
\def\eg{{\em e.g.}}
\def\etc{{\em etc.}}
\clearpage
{\em{Introduction.}}  In Kaluza--Klein theories of the ordinary type
spacetime is assumed to be of the form $M_4 \times K$; $M_4$ a
four-dimensional manifold and $K$ a compact manifold of ``internal''
co-ordinates.  Classically particles are free to move around on $K$,
thus $K$ must be assumed to be very small.  The alternative we wish to
begin to explore in this note is to somehow arrange for an absence of
translational invariance in the extra ``internal'' directions, and
then use gravity to trap particles near a four-dimensional
submanifold.  Suggestions along similar lines (without gravity) have
been made by Rubakov and Shaposhnikov [1].  For simplicity we consider
the case of five dimensions, though the general procedure immediately
generalizes to arbitrary dimensionality.  We first give a general
discussion that does not specify the particular form of the matter
giving rise to the assumed metric, and only later do we exhibit a
particular example.  The particular example we discuss also exhibits a
nice mechanism for generating a zero effective cosmological constant
from a nonzero five-dimensional cosmological constant.

{\em{The Metric.}} Let the co-ordinates be $X^A = (t,x,y,z,\xi)$;
$A=0,1,2,3,5$.  Assume that the metric is of the form
\[ 
g = -e^{2\phi(\xi)}\;(\d t)^2 + \d\x \cdot \d\x + (\d\xi)^2.
\] 
Without even having to look at field equations it is immediately
apparent that there are four Killing vectors and, therefore, four
constants of the motion for particles following geodesics
\[
E  
=  - \left(P,{\partial\over\partial t}\right)  
=  - P^A \; g_{AB} \; \left({\partial\over\partial t}\right)^B  
=  P^0\; e^{2\phi},
\]
\[
p^i  
=  \left(P,{\partial\over\partial x^i}\right) 
=  P^A \; g_{AB} \; \left({\partial\over\partial x^i}\right)^B  
=  P^i \qquad [i=1,2,3].
\]
A particle having a definite rest mass in the five dimensional sense
now satisfies
\[
-(M_5)^2 =  P^A \; g_{AB} \; P^B
=  - e^{2\phi} (P^0)^2   + \p \cdot  \p  + (P^5)^2.
\]
Consequently
\[
P^5 = \sqrt{E^2 \; e^{-2\phi} - (M_5)^2 - \p^2}
\]
with $(E,\p)$ constants of the motion.  Thus if 
\[
E < \sqrt{(M_5)^2 + \p^2} \; \; \hbox{sup}(e^\phi)
\]
any classical particle will be bound by the potential $\phi(\xi)$.
The experimental nonobservation of any extra dimensions implies that
$e^\phi$ must rise {\em very} rapidly, presumably with a length scale
of order the Planck mass.
 
To see how this gravitational trapping works quantum mechanically
consider the Klein--Gordon equation in such a background metric:
\[ 
{1\over\sqrt{-g}} \; 
\partial_A( \sqrt{-g} \; g^{AB} \; \partial_B \Psi )  
= (M_5)^2 \; \Psi.
\]
Substituting the explicit form of the metric
\[
\left[
-e^{-2\phi}\left({\partial\over\partial t}\right)^2 + 
\left({\partial\over\partial \x}\right)^2 + 
\left({\partial\over\partial\xi}\right)^2 
\right] \; \Psi 
+
\left({\partial\phi\over\partial\xi}\right) 
{\partial\Psi\over\partial\xi} = (M_5)^2 \Psi.
\] 
Try a separation of variables
\[
\Psi = \exp[-i(\omega t-\vec k \cdot \vec x)] \; 
e^{-\phi/2} \; \Psi_\perp(\xi).
\]
Then $\Psi_\perp(\xi)$ satisfies the Schrodinger equation
\[
\left[
-{1\over2}\left({\partial\over\partial\xi}\right)^2 + 
{1\over2}\left(
{1\over2} \; \phi'' + {1\over4} \; \phi' \phi' -
\omega^2 \; e^{-2\phi}
\right) 
\right] \Psi_\perp
=
-{1\over2} \left[ (M_5)^2 + \k^2 \right] \Psi_\perp.
\] 
Rewrite the eigenvalue problem as
\[ 
\left[
-{1\over2}\left({\partial\over\partial\xi}\right)^2 
+ V(\omega,\xi)
\right] \eta_n(\omega,\xi) =
- \lambda_n(\omega) \; \eta_n(\omega,\xi),
\]
with
\[ 
V(\omega,\xi)  =  
{1\over4} \; \phi''  
+ {1\over8}  \; \phi' \phi' 
- {1\over2} \; \omega^2  \; e^{-2\phi}.
\] 
Then the solutions of the Klein--Gordon equation are travelling waves
in the four usual directions but bound states in the fifth co-ordinate
\[ 
\Psi(t,\x,\xi)  =  
\exp[i(\omega t-\k\cdot\x)] \; e^{-\phi/2} \;  \eta_n(\omega,\xi).
\] 
The excitation spectrum is found by solving
\[ 
2 \lambda_n(\omega)  =  (M_5)^2   +  (\k)^2 .
\] 
Note that by choosing an appropriate normalization for $t$ we can get
\[ 
\hbox{inf}(\phi) = 0.  
\]
Then $V(\omega,\xi)$ is bounded by
\[ 
V(\omega=0,\xi) \; \geq \; 
V(\omega,\xi) \; \geq \;
V(\omega=0,\xi) - {1\over2} \; \omega^2.
\] 
Leading to the inequality
\[ 
\lambda_n(\omega) \; \leq \; \lambda_0(\omega=0)  + {1\over2} \; \omega^2.
\]
In the usual type of Kaluza--Klein model one obtains an infinite tower
of excited states of ever increasing rest mass.  For this exotic class
of Kaluza--Klein models the spectrum is significantly more complex.
Let $\lambda_n^{-1}(x)$ be the inverse function to $x = \lambda_n
(\omega)$, then
\[ 
\omega_n(\k)  = \lambda_n^{-1}\left(\half\left[ (M_5)^2  + \k^2 \right]\right).
\]
In particular the spectrum will not in general be Lorentz invariant
(\ie, will not have a $O(1,3)$ symmetry).  This should not be
surprising since our ansatz for the metric did not possess an $O(1,3)$
invariance.  We would however like to see $O(1,3)$ invariance show up
in some approximation.  To achieve this let us {\em define} $\Omega_n$
by
\[ 
2 \; \lambda_n(\Omega_n)  =  (M_5)^2; \qquad (\hbox{\ie,\ } \k  = 0).
\]
Now evaluate $\lambda_n(\omega)$ near $\omega=\Omega_n$ by using
first-order perturbation theory in $\omega^2$; we see
\[
\lambda_n(\omega) = 
\lambda_n(\Omega_n) +  
{1\over2} c_n^{-2} \; (\omega^2  - \Omega_n^2) +  
\dots,
\]
\[
c_n^{-2}  =  
\left\langle \eta_n(\Omega_n,\xi) \left|
e^{-2\phi(\xi)}
\right| \eta_n(\Omega_n,\xi )\right\rangle.
\] 
The spectrum may now be approximated by solving
\[ 
(M_5)^2 + c_n^{-2}(\omega^2 - \Omega_n^2)  + \dots =  
(M_5)^2 +  \k^2 .
\] 
Then
\[ 
\omega^2 =   \Omega_n^2  +  c_n^2 \; \k^2   +  \dots
\]
Thus in the region where first order perturbation theory is valid
[this region depends on the precise form of $\phi(\xi)$] we obtain an
approximately Lorentz invariant spectrum with rest energy $\Omega_n$
and ``effective speed of light'' $c_n$.  (In General Relativity it is
rather common to interpret $c = \sqrt{g_{00}}$ as the ``local speed of
light'', observe that $c_n^{-2} = \langle c(\xi)^{-2} \rangle$ is just
the expectation value of this local speed of light for the $n$'th
excited mode).
 
Up to this point we have nowhere used the Einstein field equations,
the discussion has accordingly been very general and qualitative.

{\em{The Field equations.}}  Starting from the metric:
\[ 
g_{00} =  -  e^{2\phi(\xi)};  
\]
\[
g_{11}  =  g_{22}  =  g_{33}  =  g_{55}  =  1;
\]
(all other components zero); it is a simple matter to compute the
Ricci tensor and Einstein tensor
\[ 
R_{0}{}^{0} =  R_{5}{}^{5}  =
  - \phi'' -  \phi' \phi'  =  - e^{-\phi} (e^\phi)'',
\]
all other components vanish;
\[ 
G^{11}  =  G^{22}  =  G^{33}  =  +  e^{-\phi} (e^\phi)'',
\] 
all other components vanish.

Consider the Einstein field equations in the form
\[ 
G_{AB}  =  + \lambda g_{AB}  +  T_{AB},
\]
we have
\begin{eqnarray*}
&1)&  T_0{}^0  =  - \lambda    =  T_5{}^5, \\
&2)&  (e^\phi)''  =  (+\lambda + T_{11}) e^\phi, \\
&3)&  T_{11}  =  T_{22}  =  T_{33};
\end{eqnarray*}
other components vanish.

So that we see that it is the pressure $p = T_{11} = T_{22} = T_{33}$
that is responsible for driving the $\xi$ dependence of $\phi$, while
the density must be independent of $\xi$.
 
In fact, a stress energy tensor of this type can easily be constructed
by considering five-dimensional Electromagnetism with a constant
electric field pointing in the $\xi$ direction.  Take:
\[ 
\L =  -\quarter \; F_{AB} \; F^{AB}.
\]
\[ 
T_{AB}  =  
-  2 {\delta\L\over\delta g^{AB}} +  g_{AB} \; \L =  
F_{AC} \; F_B{}^C  -  \quarter \; g_{AB} \; (F_{CD} \; F^{CD}).
\]
Pick
\[ 
A_0=a(\xi), \qquad A_1 = A_2 = A_3 = A_5 = 0.
\]
Then
\[ 
F_{05} = - F_{50}  =  a',
\qquad 
F_{AB} \; F^{AB}  =  -2 \; e^{-2\phi} \; (a')^2.
\]
For convenience define\footnote{Note added: Here and henceforth $E$ is
the electric field in the 5 direction, {\em not} the particle energy.}
\[ 
E  =  e^{-\phi} \; a'.
\]
Now
\begin{eqnarray*}
T_0{}^0  &=&  F_{05} F^{05} - \quarter(-2E^2) \\
         &=&  a' \, (-e^{-2\phi} \, a')  + \half E^2  \\
         &=&  - \half E^2.
\end{eqnarray*}
Similarly
\[
T_5{}^5 = -\half \; E^2; 
\qquad
T_1{}^1 = T_2{}^2 = T_3{}^3 = +\half \; E^2.
\]
The field equations now yield
\[ 
\lambda =  +\half E^2,
\qquad
(e^\phi)''  =  E^2 \; (e^\phi),
\]
with solution
\[ 
e^\phi = \cosh(E\xi),
\qquad
a = \sinh(E\xi).
\]
Note that the equation of motion for $F$ is
\begin{eqnarray*}
{1\over\sqrt{-g}} \partial_A( \sqrt{-g} \; F_{AB})  =  0     
&\to& \partial_5(e^\phi[-e^{-2\phi} \; a'])  =  0 \\
&\to& \partial_5(e^{-\phi} \; a')  =  0 \\
&\to& \partial_5(E)  =  0,
\end{eqnarray*}
consistent with the result $\lambda = +\half \; E^2$.
 
What has happened in this model is that a five-dimensional
cosmological constant has conspired with a constant electric field $E=
\sqrt{2\lambda}$, to yield a four-dimensional submanifold of zero
four-dimensional cosmological constant, together with a potential
$\phi = \ln \cosh(E\xi)$ that traps particles in the fifth direction.
The quantum mechanical trapping is in this case governed by the
Schrodinger equation
\[
\left[ 
-\half \left({\partial\over\partial\xi}\right)^2 
+ {1\over8} E^2 
-\left(\half \omega^2 - {1\over8} E^2\right) \sech^2(E\phi) 
\right] \eta_n
=
\lambda_n \; \eta_n.
\]
This potential is the well-known Rosen--Morse potential, the exact
eigenvalues are known to be [2]
\[ 
\lambda_n = 
\half \left\{ \left(\omega - \left[n + \half\right] E\right)^2  - 
\left({E\over2}\right)^2 \right\}.
\]
We solve for the exact spectrum
\[ 
\omega_n(\k) =  
\left[n + \half\right] E  
\pm \left[\left({E\over2}\right)^2  + (M_5)^2  + (\k)^2 \right]^{1/2}.
\]
We define a 4-dimensional mass by
\[
(m_4)^2 = (M_5)^2 + \left({E\over2}\right)^2.
\]
Then
\[
\omega_n(\k)  =  \left[n + \half\right]E  
\pm \left[(m_4)^2 + \k^2\right]^{1/2}.
\]
We see that apart from a momentum independent shift the spectrum is in
this case exactly Lorentzian.  The appearance of two branches is not
alarming.  These are just the usual particle antiparticle branches.
It should be emphasized that it is quite possible to keep $m_4$ small
(compared to $E$).  This requires that the five dimensional mass term
be tachyonic.  (\eg, $(M_5)^2 = - E^2/4$ for massless matter).  For some
closing comments, note that there is no need for the five dimensional
``Electromagnetism'' considered here to have anything to do with
ordinary electromagnetism.  In addition, the total action of the field
configuration considered here is zero.
\[ 
S = 
\int \d^5x  \; \sqrt{-g}  \; (R + 2\lambda  + 2\L )
=  
\int \d^5x \; \sqrt{-g} \; (-2E^2  + E^2  + E^2 )
=  0.
\]
Thus we expect field configurations of this type to be important for
five dimensional quantum gravity even though they are not
asymptotically flat.

{\em{Conclusion.}}  We have looked at a general class of exotic
Kaluza--Klein models in which the particles of the observable world
are gravitationally trapped on a four dimensional submanifold of the
``real'' world.  We have seen that it is generally possible to get a
low energy spectrum that approximately respects Lorentz invariance,
even though the metric ansatz does not possess an $O(1,3)$ symmetry.
Finally we have exhibited a particular model in five dimensions that
illustrates the discussed phenomena.  The five dimensional model is
interesting in other respects.  It exhibits an interesting mechanism
for cancelling a five dimensional cosmological constant.  In addition
it exhibits a mechanism for taking what would be a tachyon in flat
five dimensional spacetime and producing an ordinary particle in the
four dimensional submanifold.
 
This work was supported by the Division of High Energy Physics of the
U.S. Department of Energy under contract DE-FG03-84ER40168.
 
\bigskip

{\em{References}}

\bigskip
 
[1] V.A. Rubakov and M.E. Shaposhnikov, 
Phys. Lett. {\bf 125B} (1983) 136.
 
[2] M.M. Nieto, Phys. Rev. {\bf A 17} (1978) 1273.

\bigskip

\hrule

\smallskip

\hrule

\bigskip

\section*{Note added 14 years later:}

Submanifold variants of Kaluza--Klein theory, in which the (3+1)
dimensional matter is somehow constrained to live on a
four-dimensional submanifold of a higher dimensional spacetime have a
very long history, going back {\em at least} as far as
Joseph~\cite{Joseph}.  Interest in this approach in a particle physics
context, using {\em non-gravitational} fields to effect the trapping,
is largely due to the early paper of Rubakov and
Shaposhnikov~\cite{Rubakov}.

After initial appearance of this present paper~\cite{Visser:1985qm},
the approach was somewhat modified by
Squires~\cite{Squires:1986aq}. Important related papers that further
develop this type of approach are those of Laguna--Castillo and
Matzner~\cite{Laguna--Castillo}, and of Gibbons and
Wiltshire~\cite{Gibbons:1987wg}. The field then lay largely fallow for
several years
\cite{Hubsch:1989hc,Maia:1993fw,Bleyer:1995wg,Maia:1995wh,Davidson:1996dm},
until the appearance of the review article by Overduin and
Wesson~\cite{Overduin:1997pn}, which compares and contrasts some of
the non-standard approaches to Kaluza--Klein gravity that have been
considered in the literature.

In the last year and a half interest has picked up considerably
\cite{Maia:1998jb,Kalbermann:1998hu,Gogberashvili:1998vx,%
Gogberashvili:1998iu,Gogberashvili:1999tb,Randall:1999vf,%
Csaki:1999jh,Gogberashvili:1999ad,Csaki:1999mz}. Note that this topic
is one for which e-prints are widely scattered, appearing on any one
of the {\sf hep-th}, {\sf hep-ph}, or {\sf gr-qc} archives.

Perhaps the best known of these recent papers
are~\cite{Randall:1999vf,Csaki:1999jh,Csaki:1999mz}, which work
specifically within a ``brane'' context arising from the low-energy
limit of a fundamental string theory.\footnote{There are many other
related papers and I am being highly selective here.} It should be
borne in mind however, that many features of the ``non-compact extra
dimension'' paradigm are generic to any low-energy effective field
theory, and are largely independent of one's views regarding the
full-fledged theory of quantum gravity.


\end{document}